\documentclass[aps,prl,twocolumn,letterpaper,nobibnotes,superscriptaddress,longbibliography,nofootinbib]{revtex4-2}
\usepackage{amssymb}
\usepackage{amsmath}
\usepackage{graphicx}
\usepackage{extarrows}
\usepackage{url}
\usepackage{varioref}
\usepackage{tabularx}
\usepackage{epsfig}
\usepackage{longtable}
\usepackage{amsfonts}
\usepackage{rotating}
\usepackage{bbold}
\usepackage{hhline}
\usepackage{braket}

\usepackage{lineno}
\usepackage{color}
\usepackage[normalem]{ulem}
\newcommand\redsout{\bgroup\markoverwith
{\textcolor{red}{\rule[.5ex]{2pt}{0.7pt}}}\ULon}
\usepackage{xcolor}
\usepackage{float}
\usepackage{verbatim}
\usepackage{amsfonts}
\usepackage{float}

\usepackage[unicode=true,bookmarks=true,bookmarksnumbered=false,bookmarksopen=false,breaklinks=false,pdfborder={0 0 1},backref=false,colorlinks=true]{hyperref}

\hypersetup{linkcolor=magenta,urlcolor=blue,citecolor=blue,pdfstartview={FitH},unicode=true}

\def\be{\begin{equation}}
\def\ee{\end{equation}}
\def\bea{\begin{eqnarray}}
\def\eea{\end{eqnarray}}

\def\tbf{\textbf}

\def\im{{\rm i}}
\def\e{{\rm e}}

\begin{document}
\title{Adiabatically compressing chiral p-wave Bose-Einstein condensates into the lowest landau level}

\author{Xinyang Yu}
\affiliation{State Key Laboratory of Surface Physics, Institute of Nanoelectronics and Quantum Computing,
and Department of Physics, Fudan University, Shanghai 200433, China
}

\author{Xingze Qiu}
\affiliation{State Key Laboratory of Surface Physics, Institute of Nanoelectronics and Quantum Computing,
and Department of Physics, Fudan University, Shanghai 200433, China
}
\affiliation{School of Physics Science and Engineering, Tongji University, Shanghai 200092, China} 

\author{Xiaopeng Li}
\email{xiaopeng\underline{ }li@fudan.edu.cn} 
\affiliation{State Key Laboratory of Surface Physics, Institute of Nanoelectronics and Quantum Computing,
and Department of Physics, Fudan University, Shanghai 200433, China
}
\affiliation{Shanghai Qi Zhi Institute, AI Tower, Xuhui District, Shanghai 200232, China} 
\affiliation{Shanghai Artificial Intelligience Laboratory, Shanghai 200232, China} 
\affiliation{Shanghai Research Center for Quantum Sciences, Shanghai 201315, China} 

\begin{abstract}

There has been much recent progress in controlling $p$-orbital degrees of freedom in optical lattices, for example with lattice shaking, sublattice swapping, and lattice potential programming. Here, we present a protocol of preparing lowest Landau level (LLL) states of cold atoms by adiabatically compressing $p$-orbital Bose-Einstein condensates confined in two-dimensional optical lattices. The system starts from a chiral $p+ip$ Bose-Einstein condensate (BEC) state, which acquires finite angular momentum by spontaneous symmetry breaking. Such chiral BEC states have been achieved in recent optical lattice experiments for cold atoms loaded in the $p$-bands. Through an adiabatic adjustment of the lattice potential, we compress the three-dimensional BEC into a two-dimensional system, in which the orbital degrees of freedom continuously morph into LLL states. This process is enforced by the discrete rotation symmetry of the lattice potential. The final quantum state inherits large angular momentum from the original chiral $p+ip$ state, with one quantized unit per particle. We investigate the quantum many-body ground state of interacting bosons in the LLL considering contact  repulsion. This leads to an exotic gapped BEC state. Our theory can be readily tested in experiments for the required techniques are all accessible to the current optical lattice experiments.

 
\end{abstract}

\date{\today}
\maketitle

{\it Introduction.---} \label{section Intro}
Landau Level describes charged  particles moving in an external magnetic field. 
It exhibits a flat-band energy spectrum with an extensive degeneracy of single-particle states, leading to effective amplification of many-body interaction effects. As a result, interacting bosons loaded in the lowest Landau level (LLL) 
could escape the conventional scenario of condensing at a single energy minimum, but instead form more exotic quantum many-body states such as Abrikosov vortex lattices and bosonic quantum Hall state (BQHE)~\cite{CooperReview,FetterReview}.

With cold atoms that are charge-neutral, synthesizing artificial gauge fields to mimic the Landau level physics has been attracting continuous research interests in the last one or two decades~\cite{DalibardRMP2011}. 

One approach is to utilize the Raman-assisted tunneling process, where the accumulated phase shift of a hopping atom emulates the Aharonov-Bohm phase of charged particles in the magnetic field~\cite{periodic17,raman11}. With this approach, the Hofstadter Hamiltonian and a two-atom analogue of the $\nu = 1/2$ bosonic Laughlin state in the interacting Hofstadter system~\cite{https://doi.org/10.48550/arxiv.2210.10919}. has been realized~\cite{hofstadter13}.
However, most of these studies are limited to few-particle systems and the generalization to a genuine many-body setting is highly nontrivial~\cite{LatticeExp1,LatticeExp2,CooperPrl2013,LatiiceExp3}.

An alternative approach to create synthetic magnetic fields is to rotate the trap potential. 
In the rotating frame, the Coriolis force experienced by the rotating BEC (rBEC) resembles the Lorentz force experienced by charged particles in a magnetic field.
With this approach, vortex lattice states have been observed about two decades ago  having a total angular momentum of $L\sim O(N)$ ($N$ is the atom number)~\cite{rBECexp1,rBECexp2}. 
Recent experiments~\cite{2021geometric,2022crystallization} have successfully prepared a LLL condensate consisting of $N=8\times 10^5$ atoms, carrying $10^3$ quanta of angular momentum per particle, for which crystallization of the condensate has been observed.

In optical lattice experiments, there has been much progress in controlling $p$-orbital degrees of freedom in optical lattices, for example with lattice shaking~\cite{2013_Chin_Shaking,2022_Schneider_NatPhys,Sun_2023}, and sublattice swapping~\cite{wirth2011evidence,jin2021evidence,wang2021evidence}. These advances led to the experimental realization of the $p$-orbital BEC ($p$-BEC), which provides novel opportunities to control atomic wavefunctions. In $p$-BEC experiments, bosonic atoms are loaded into  the $p$-orbital bands of the optical lattices. These atoms are found to condense  at one specific high symmetry point of the $p$-bands, and a chiral $p_x+ip_y$ BEC has been observed~\cite{wirth2011evidence,wang2021evidence}. This spontaneously breaks time-reversal symmetry (TRS) and results in $p$-BEC states belonging to non-trivial representations of the discrete lattice rotation operations. 

\begin{figure*}[htp]
\includegraphics[width=\linewidth]{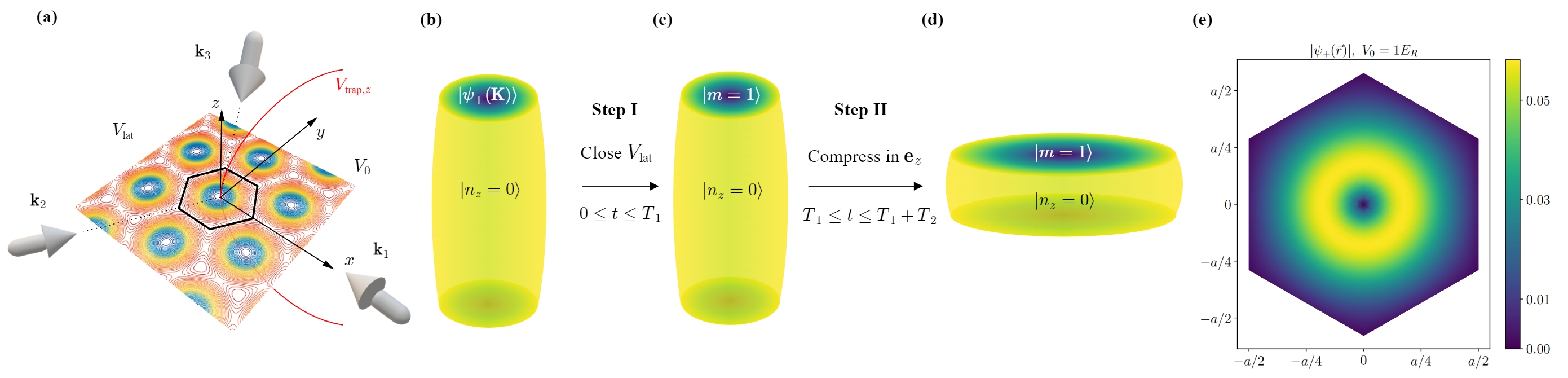}
\caption{Schematic illustration of our adiabatic compress protocol with a specific triangular optical lattice example.
(\tbf{a}) Potential field at $t=0$ (Eq.~\eqref{eqn:potential0}). In the $xy$-plane, a triangular optical lattice $V_{\text{lat}}$ is generated by three $120^{\circ}$ intersecting lasers and color represents the value of $V_{\text{lat}}$. The central black hexagon is a Wigner-Seitz unit cell of the triangular lattice, which is the only unit cell taken into account in \tbf{Step I}. In the $\mathbf{e}_z$ direction, a harmonic trap $V_{\text{trap,z}}$ is imposed to confine atoms.
(\tbf{b}-\tbf{d}) Two steps of our protocol, with cylinders denoting the contour of the single-body wave function $\psi(\mathbf{r},t)$ in the real space and color indicating its absolute value. The $z$ component of $|\psi(t)\rangle$ is always in the ground state $|n_z=0\rangle$ of $V_{\text{trap},z}$.
(\tbf{b}) The initial $xy$ component at $t=0$ is prepared as the $p$-BEC state $|\psi_+\rangle$.
In \textbf{Step I}, we turn off the optical lattice and simultaneously open a harmonic trap in the $xy$-plane (Eq.~\eqref{eqn:potentialstepI}). 
(\tbf{c}) After step I, the $xy$ component becomes our target LLL state $|m=1\rangle$. $\psi(T_1)$ still has a 3d tube shape. 
In \textbf{Step II}, we adjust the trap frequency $\omega_{xy}$ and $\omega_z$ to compress the system in the $\mathbf{e}_z$ direction (Eq.~\eqref{eqn:potentialstepII}).
(\tbf{d}) After step II, $\psi(T_1+T_2)$ is compressed to a quasi-2d plate shape.
(\tbf{e}) Real space wavefunction of the initial $xy$ component $|\psi_{+}(\mathbf{r})|$ in the central unit cell with $V_0=1E_R$.} 

\label{fig:illustratition}
\end{figure*}

In this paper, we propose a novel adiabatic compressing approach to prepare LLL states from chiral $p+ip$ BEC states.  Through symmetry analysis and  numerical simulations, we demonstrate how the adiabatic compressing transforms the initial $p$-BEC states into LLL states with precise angular momentum $L=N$.  
We analyze the quantum many-body ground state of the interacting bosons in LLL, and find an exotic gapped BEC state without a gapless Goldstone mode, drastically different from conventional BEC. We provide analytic expressions for the first excited state and its excitation energy, which are validated by numerical simulations. 
The gapped BEC state largely relies on the angular-momentum being precisely $L = N$, which holds naturally in our adiabatic compressing protocol. Projecting the quantum state into the Hilbert space sector of $L=N$ is challenging with other approaches of preparing LLL states for example with Raman schemes or rotating the BEC.

{\it Adiabatic compressing $p$-wave BEC states.---}\label{subsec:$p$-BEC}
We consider a chiral $p_x +i p_y$ BEC 
 ($p$-BEC)
state trapped by a 2d optical lattice potential $V_{\text{lat}}$ in the $xy$ plane and a harmonic potential $V_{\text{trap},z}$ along the $z$ direction (Fig.~\ref{fig:illustratition}(\tbf{a})), described by~\cite{leggett2001bose}:
\begin{equation}\label{eqn:hamiltonian}
    \hat{H} = \sum_{i=1}^{N} \left[\frac{\hat{p_i}^2}{2M} + \hat{V}(\mathbf{r},t) \right] + 2\pi g\sum_{i<j} \delta(\mathbf{r}_i-\mathbf{r}_j)
\end{equation}
with $M$ the atomic mass, $\hat{V}$ the full trap potential, $g=2\hbar^2a_s/M$ the interaction strength and $a_s$ the s-wave scattering length which can be tuned with Feshbach Resonance techniques. 

We assume the band minima of the second Bloch energy band occur at the high symmetry points of the Brillouin Zone (BZ), denoted as $\pm \mathbf{K}$. The chiral $p$-BEC state $\psi_{+}(\mathbf{K})$ spontaneously breaks TRS and condense at one particular high symmetry point, 
say $\mathbf{K}$.
It can be expanded as $|\psi_{+}(\mathbf{K})\rangle = \sum_{\mathbf{R}} \exp (i\mathbf{K}\cdot\mathbf{R}) |\omega_{+},\mathbf{R}\rangle$ where $|\omega_{+},\mathbf{R}\rangle$ is the Wannier state of $p_{+} = p_x + ip_y$ orbital localized at site $\mathbf{R}$ (Its time reversal counterpart $p_{-}= p_x - i p_y$ for $\omega_{-}$ and $\psi_{-}$ is equivalent in further analysis), which has one unit of local angular momentum. Remarkably, our adiabatic protocol focuses solely on the contribution from the central unit cell within the entire optical lattice(Fig.~\ref{fig:illustratition}).

\begin{figure}[h]
\includegraphics[width=\linewidth]{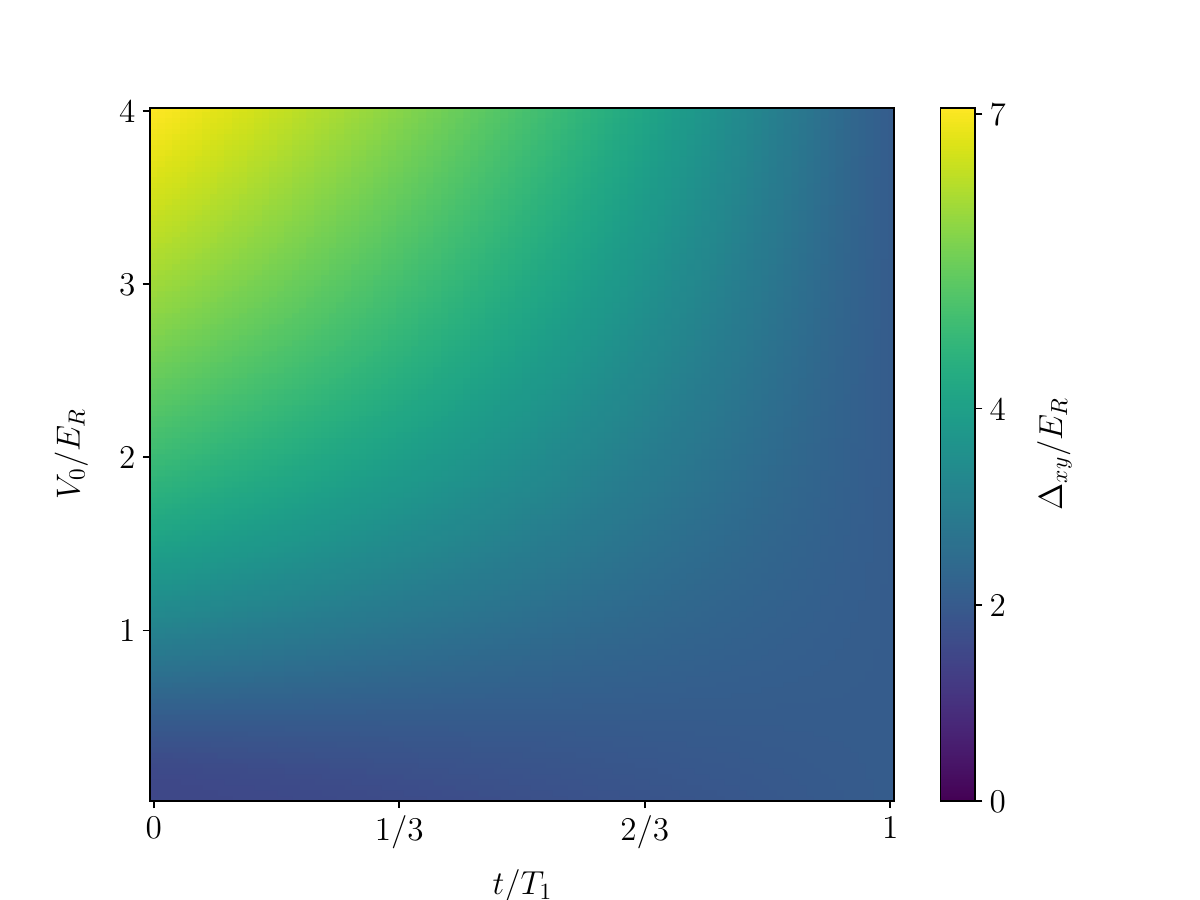}
\caption{Gap of the $xy$ sector $\Delta_{xy}$ of the non-interacting hamiltonian $\hat{H}_{xy}$ in the symmetrical subspace $\mathcal{H}_{\text{sym}}$ with $V_0$ takes value in the range $(0,4E_R)$.$\ \omega_{xy} \equiv 1$ ($\omega_{xy,z}$ is in units of $\hbar k_L^2/M$) in this step. $\Delta_{xy}(t)$ has an exact lower bound $\Delta_{xy}(T_1)=2\hbar\omega_{xy}$.
The existence of finite gaps ensures the adiabaticity of our adiabatic protocol. Here we set $T_1=10\hbar/E_R$, and use linear adiabatic path between initial and final hamiltonian.
}
\label{fig:energy_level}
\end{figure}

The main feature of $p$-BEC state $\psi_{+}(\mathbf{K})$ is its transformation property under lattice rotation operations. The little group of high symmetry point $\mathbf{K}$ is a discrete rotation group generated by $\hat{R}=\exp(-\im \theta\hat{L}_z/\hbar)$. Under this discrete rotation operation, $p$-BEC state transforms as (proved in Supplementary Material)

\begin{equation}\label{eqn:sym}
     \hat{R}|\psi_{+}(\mathbf{K})\rangle = \exp(-\im \theta)|\psi_{+}(\mathbf{K})\rangle.
\end{equation}
Based on this symmetry property, we devise a two-step adiabatic protocol  (Fig.~\ref{fig:illustratition}(\tbf{b})-(\tbf{d})) to prepare quasi-2d LLL states with global angular momentum $L=N$ by compressing the initial 3d tubular $p$-BEC state.
The symmetry $\hat{R}(t)=\exp(-\im \theta\hat{L}_z/\hbar) \equiv \exp(-\im \theta)$ ensures the adiabaticity of the protocol(Fig.~\ref{fig:energy_level}) and guarantees that each atom carries one unit of global angular momentum $\hat{L}_z/\hbar=1$ in the final stage when $\hat{L}_z$ becomes a good quantum number. 

{\it Target LLL states.---}\label{subsec:LLL}
At the end of our adiabatic protocol, the single body hamiltonian becomes 2D isotropic harmonic oscillator hamiltonian $\hat{H}_{\text{har}}=\hat{p}^2/2M + M\omega_{xy}^2(\hat{x}^2+\hat{y}^2)/2$. It possesses continuous SO(2) rotation symmetry and can be diagonalized simultaneously with $\hat{L}_z$ in the $|n,m\rangle$ basis as $\hat{H}_{\text{har}}|n,m\rangle = n\hbar\omega_{xy} |n,m\rangle $, $\hat{L}_{z}|n,m\rangle = m\hbar |n,m\rangle$, where $n\in \mathbb{N}$ and $m=-n,-n+2,...,n-2,n$ are the energy and angular momentum quantum numbers. The states with $n=m$ are exactly the same as the LLL states of charged particles in the magnetic field under the symmetric  gauge~\cite{tong2016quantum} with wavefunctions $\phi_m(z) = \langle z |m\rangle = \frac{z^m}{\sqrt{\pi m! }l}e^{-|z|^2/2}$, where dimensionless complex coordinate $z=l^{-1}(x+iy)$, length scale $l=\sqrt{\hbar/M\omega_{xy}}$, and $|m\rangle:=|n,m=n\rangle$. 

Our target state of the adiabatic preparation process is $|m=1\rangle$ in the LLL, which carries one unit of global angular momentum per particle. It shares the same symmetry as the $p$-BEC state $|\psi_{+}(\mathbf{K})\rangle$, characterized by $\hat{R}|m=1\rangle = \exp(-i\theta) |m=1\rangle$ under discrete lattice rotations.

\begin{figure}[h]
\includegraphics[width=\linewidth]{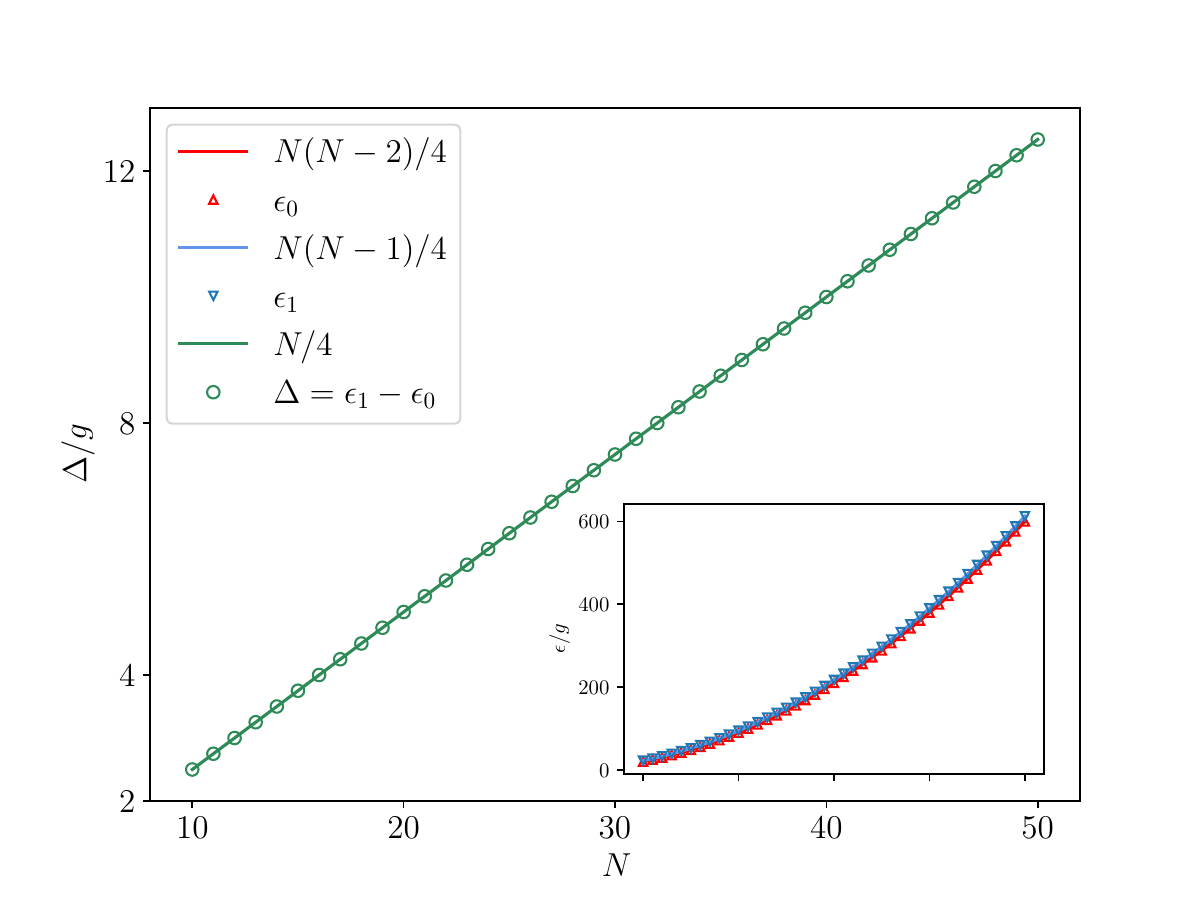}
\caption{Ground state energy $\epsilon_0$, the first excited state energy $\epsilon_1$ and the gap $\Delta = \epsilon_1 - \epsilon_0$ of $\hat{H}$ on $\mathcal{H}_{N,L=N}$ as a function of particle number $N$. Points correspond to values obtained by exact diagonalization of the many-body hamiltonian $\hat{H}$ given by Eq.~\eqref{eqn:hamiltonian}, while solid lines to their corresponding (conjectured) analytical value. $\epsilon_0$ has a proved analytical value $\epsilon_0(N) = gN(N-2)/4$, while we conjecture that $\epsilon_1$ has an analytical expression $\epsilon_1(N) = gN(N-1)/4$, and accordingly the gap $\Delta$ has a conjectured analytical value $\Delta(N)=N/4$.} 
\label{fig:gap}
\end{figure}

{\it Gapped ground state in the LLL.---}\label{sec:GS} 
We set $g=0$ in Eq.~\eqref{eqn:hamiltonian} during the adiabatic compress process and achieve the quasi-2D target LLL condensed state $|m=1,n_z=0\rangle^{\otimes N}$ with a total angular momentum $L=N$. We now investigate the effect of a small repulsive interaction $g>0$ on the system. 

When the interaction strength $g \ll \hbar \omega_{xy,z}$ , LLL states will not be scattered to excitation states $n_z>0$ along the $z$ direction~\cite{CooperReview,LiebPRA2009}, nor to higher LL states (Supplementary Materials). Hence, we can focus on $\mathcal{H}_{N,L}$ consisting of only 2d LLL states with fixed particle number $N$ and angular momentum $L$. The projection of many-body hamiltonian $\hat{H}$ to the LLL states $\mathcal{H}_{N,L}$ is:
\begin{equation}\label{eqn:SecondQuantizedH}
    \hat{H} = \hat{H}_0 + \hat{V}_{\text{int}} = \omega(\hat{L}+\hat{N}) + \sum_{mnkl}u_{mnkl}\hat{a}_m^{\dagger} \hat{a}_n^{\dagger} \hat{a}_k \hat{a}_l
\end{equation}
where $u_{mnkl}=2^{-(m+n+1)}\ \frac{(m+n)!}{\sqrt{m!n!k!l!}}g \ \delta_{m+n,k+l}$ and $\hat{a}_m$ represents the annihilation operator corresponding to $|m\rangle$.

Noticing that the single-particle hamiltonian can be expressed as $\omega(\hat{L}+\hat{N})$, indicating that all states in $\mathcal{H}_{N,L}$ are exactly degenerate with $\hat{H_0}$ and constitute the flat LLL in our system. Properties such as ground state or first excited state are solely determined by the interaction term $\hat{V}_{\text{int}}$.

We employ the Exact Diagonalization (ED) method to evaluate the ground state $\psi_0$ and its energy $\epsilon_0$, the first excited state $\psi_1$ and its energy $\epsilon_1$, as well as the gap $\Delta = \epsilon_1 - \epsilon_0$ in the subspace $\mathcal{H}_{N,L=N}$. The energy dependencies on $N$ are depicted in Fig.~\ref{fig:gap}. We find that (for $N\geq 10$) the energies are given precisely by $\epsilon_0 = gN(N-2)/4$, $\epsilon_1 = gN(N-1)/4$ and $\Delta = gN/4$, with arbitrary numerical precision. In terms of the center of mass coordinate $z_c = \sum_{i=1}^{N}z_i/N$, the numerically obtained ground state $\psi_{0}(z_1,...,z_N)=\prod_{i=1}^{N}\left [(z_i-z_c)\cdot\e^{-|z_i|^2/2}\right]$ recovers previous theoretical results~\cite{GSexpression,lewin2009strongly}. 
Its one-body reduced density matrix $\hat{\rho}_1 =\Big(1 - \frac{2}{N}\Big)|1\rangle\langle1| + \frac{1}{N}\Big(|0\rangle\langle0|+|2\rangle\langle2|\Big) + O(\frac{1}{N^2})$ and hence $\psi_0$ is fully condensed at $|1\rangle$ in the thermodynamic limit~\cite{ODLRO,DoAttractive}, spontaneously breaking the global $U(1)$ phase symmetry $a_m \rightarrow a_m \exp(\im \varphi)$.
The first excited state is calculated as:

\begin{equation}
     \psi_{1} = z_c \left[\sum_{1 \leq p_1 < ... < p_{N-1} \leq N} \prod_{i=1}^{N-1} (z_{p_i}-z_c) \right]\e^{-\sum_{i]1}^{N}|z_i|^2/2}
\end{equation}

We theoretically demonstrate that $\psi_{1}$ is an exact eigenstate of $\hat{V}_{\text{int}}$ with eigenenergy $\epsilon_1$ (Supplementary material). However we cannot establish analytically whether $\psi_1$ corresponds to the first excited state, i.e., there might still exist other eigenstates of $\hat{V}_{\text{int}}$ with energy $\epsilon \in (\epsilon_0, \epsilon_1)$.

Based on the aforementioned results, we put forward the conjecture that $\hat{H}$ exhibits a gap in the subspace $\mathcal{H}_{N,L=N}$ with a gap size $\Delta = gN/4$ and corresponding first excited state energy $\epsilon_1 = gN(N-1)/4$ in the thermodynamic limit. As a consequence, our system will provide an example of spontaneous breaking of the global continuous symmetry without a gapless Goldstone mode, which contrasts with the conventional homogeneous or harmonically-trapped BEC~\cite{BECPitaevskii}. As Goldstone’s theorem applies solely to hamiltonians that preserve some translational symmetries~\cite{SSBIntro}, the violation of Goldstone theorem in our system can be traced back to the fact that our LLL states are in the symmetric gauge which break all kinds of continuous and discrete translational symmetry.


{\it Numerical demonstration of adiabaticity.---}\label{subsec:adiabatic}
Our adiabatic compress protocol (Fig.~\ref{fig:illustratition}) applies to chiral $p$-BEC state condensed at high symmetry point in optical lattice with arbitrary shape. Without loss of generality, we now demonstrate in detail the adiabaticity of our protocol with a specific triangular optical lattice example based on two recent experiments~\cite{jin2021evidence,wang2021evidence}.

We consider a triangular optical lattice (Fig.~\ref{fig:illustratition}(\tbf{a})), described by the potential $V_{\text{lat}} (\mathbf{r}) = -2V_0\sum_{i=1}^{3}\cos(\mathbf{k}_i\cdot \mathbf{r})$ with $\mathbf{k}_1 = (-1,0)k_L,\ \mathbf{k}_{2,3} = (1/2,\pm\sqrt{3}/2)k_L$, where $V_0$ is the lattice potential depth and $k_L$ is the wavelength of the lasers used to generate optical lattice. The relevant energy scale is the single-photon recoil energy $E_R=\hbar^2k_L^2/2M$.

The chiral $p$-BEC state corresponds to our triangular lattice is $\psi_{+}(\mathbf{K})$ (Supplementary Material). The little group of the high symmetry point $\mathbf{K}$ is $C_6$ generated by rotation $\hat{R}=\exp(-\im \theta\hat{L}_z/\hbar)$ with $\theta=\pi/3$. 

{\bf At} $t=0$, we prepare the system in the $p$-BEC state $|\psi_+(\mathbf{K})\rangle$ of the triangular lattice $V_{\text{lat}}$ with only one unit cell in the $xy$-plane (Fig.~\ref{fig:illustratition}(\tbf{e})) and the ground state $|n_z=0\rangle$ of the harmonic trap $V_{\text{trap},z}=1/2M\omega_z^2z^2$ in the z direction. The initial potential (Fig.~\ref{fig:illustratition}(\tbf{a})) is given by
\begin{align}
    V(\mathbf{r},t=0) &=  V_{\text{lat}} + V_{\text{trap},z} \notag \\
    &= -2V_0\sum_{i=1}^{3}\cos(\mathbf{k}_i\cdot \mathbf{r}) + \frac{1}{2}M\omega_z^2z^2
\end{align}

The initial single body state is a product state of its $xy$ component and $z$ component, given by $|\psi(t=0)\rangle = |\psi_{+}(\mathbf{K}), n_z=0\rangle $. 
It has a tubular shape in 3D real space, as shown in Fig.~\ref{fig:illustratition}(\textbf{b}). The choice of $V_0$ and $\omega_z$ only affects the radius and length of the tube. 

During the adiabatic evolution, we turn off the interaction between atoms and consider only the single-body hamiltonian $\hat{H}_i(t)=\hat{p_i}^2/2M + \hat{V}(\mathbf{r},t) $. As a result, the $xy$ and $z$ sector decouple and the many-body state of the system is always a product state $(|\psi_{xy}(t), \psi_z(t)\rangle)^{\otimes N}$.

The final potential of the first step at $t=T_1$ is a 3D harmonic trap containing the 2D isotropic harmonic potential in the $xy$-plane:
\begin{align}\label{eqn:potential0}
    V(\mathbf{r},t=T) &=  V_{\text{trap},xy} + V_{\text{trap},z} \notag \\
    &= \frac{1}{2}M\omega_{xy}^2(x^2+y^2) + \frac{1}{2}M\omega_z^2z^2
\end{align}

In the first step (Fig.~\ref{fig:illustratition}(\tbf{b}-\tbf{c})), we use a simple linear adiabatic path to connect the initial and final potential, which corresponds to adiabatically decrease the depth of the optical lattice $V_\text{lat}$ and increase the strength of the harmonic trap $V_{\text{trap},xy}$ in the $xy$ plane. 
\begin{equation}\label{eqn:potentialstepI}
    V(\mathbf{r},t) = (1-\frac{t}{T_1})V_{\text{lat}} + \frac{t}{T_1} V_{\text{trap,xy}} + V_{\text{trap,z}} \ , \ 0\leq t\leq T_1
\end{equation}

Because that $\hat{R}$ represents the lattice rotation operation of triangular lattice and $V_{\text{trap},xy}$ has full SO(2) rotation symmetry, $[\hat{R},V_\text{lat}] = [\hat{R},V_{\text{trap},xy}] = 0$. As a result, $[\hat{R},\hat{H}_i(t)] = 0$ and $|\psi(t)\rangle$ is always a eigenstate of $\hat{R}$ during the adiabatic evolution process:
\begin{equation}
    \hat{R}|\psi(t)\rangle = \exp(-\im \theta)|\psi(t)\rangle  \  ,\ \theta = \pi/3
\end{equation}

As a result, we can restrict ourselves to the subspace $\mathcal{H}_{\text{sym}} = \left\{ |\psi\rangle \big| \ \hat{R}|\psi\rangle = \exp(-\im \theta)|\psi\rangle\right\}$, where $|\psi(0)\rangle$ and $|\psi(T_1)\rangle = |m=1, n_z=0\rangle$ are the ground states of $\hat{H}_{i}(0)$ and $\hat{H}_{i}(T_1)$ respectively. 

Numerical calculation of the gap $\Delta(t)$ of $\hat{H}_i(t)$ in the first step is shown in Fig.~\ref{fig:energy_level} for various optical lattice depths $V_0$. It is observed that the energy gap opens up over a large range of lattice depths from the shallow lattice case $V_0 \rightarrow 0$ to the deep lattice case $V_0 = 4E_R$. Therefore, the adiabatic theorem implies that we can reach the target state $|\psi(T_1)\rangle$ with a evolution time $T_1=O  (1/\underset{0\leq t\leq T_1}{\min}\Delta(t)) = O((2\hbar \omega_{xy}(T_1))^{-1})$.

In the second step (Fig.~\ref{fig:illustratition}(\tbf{c}-\tbf{d})), we adiabatically decrease $\omega_{xy}$ and increase $\omega_z$ to compress the state. 
The potential in the second step is
\begin{equation}\label{eqn:potentialstepII}
    V(\mathbf{r},t) = \frac{1}{2}M \omega_{xy}^2(t) (x^2+y^2) + \frac{1}{2}M \omega_z^2(t) z^2 \ ,\ 0\leq t-T_1\leq T_2
\end{equation}

$\hat{R}\equiv\exp(-\im \theta)$ is still conserved and the gap $\Delta(t) = \min(2\omega_{xy}(t),2\omega_z(t))$ (Supplementary Material) remains finite as long as both traps in $xy$ plane and $z$ direction are open. Therefore, arbitrary adiabatic path to adjust the trap frequency is admissible to compress the system. 

The result of our whole adiabatic evolution is a flattened quasi 2D state $|\psi(T_1+T_2)\rangle = |m=1, n_z=0\rangle$ whose $xy$ component is a LLL state. Compared to the fact that conventional rBEC method is sensitive to fluctuations in both the rotating angular velocity and trap geometry, our adiabatic protocol is robust against experimental parameters $V_0,\omega_{xy},\omega_{z}$.

{\it Conclusion.---} \label{sec:Conclusion}
In this work, we present a novel adiabatic scheme for preparing bosonic LLL states without the need for synthetic magnetic fields. Our approach leverages the symmetry properties of experimentally accessible $p$-BEC states to build up non-zero global angular momentum $L=N$ within the mean-field quantum Hall regime $L \sim O(N)$. Existence of a finite gap in the adiabatic evolution ensures the feasibility of our method. Moreover, the gapped nature of the protocol remains robust against variations in parameters such as $V_0,\omega_{xy},\omega_{z}$ enabling experimentally accessible implementation. When a repulsive interaction is turned on, the ground state of our system becomes a exotic gapped BEC state where Goldstone theorem does not hold.
We believe our proposed protocol is a compelling candidate for further experimental investigation.

{\it Acknowledgement.---}
We acknowledge helpful discussion with Andreas Hemmerich, W. Vincent Liu, and Xiaoji Zhou.  
This work is supported by National Program on Key Basic Research Project of China (Grant
No.\ 2021YFA1400900), National Natural Science Foundation of China (Grants No.\ 11934002), Shanghai Municipal Science and Technology Major Project (Grant No.\ 2019SHZDZX01), Shanghai Science Foundation(Grants No.\ 21QA1400500), and Shanghai Qi Zhi Institute.

\bibliography{references}

\onecolumngrid


\newpage
\renewcommand{\theequation}{S\arabic{equation}}
\renewcommand{\thesection}{S-\arabic{section}}
\renewcommand{\thefigure}{S\arabic{figure}}
\renewcommand{\thetable}{S\arabic{table}}
\setcounter{equation}{0}
\setcounter{figure}{0}
\setcounter{table}{0}

\setlength{\arrayrulewidth}{0.4mm}
\setlength{\tabcolsep}{15pt}
\renewcommand{\arraystretch}{1.8}
\newcolumntype{M}[1]{>{\centering\arraybackslash}m{#1}}

\begin{center}
    \Huge Supplementary Material
\end{center}

\section{S1. Rotational symmetry of $p$-BEC states}
\label{sec:appendixA}
In this section, we will utilize the fact that the $p_x + ip_y$ Wannier state localized at lattice site $\mathbf{a}$, denoted as $|\omega_{+},\mathbf{a}\rangle$, has one unit of local angular momentum~\cite{wang2021evidence} to prove Eq.~\eqref{eqn:sym}. This fact can be expressed mathematically as $\hat{L}_{z,\mathbf{a}}|\omega_{+},\mathbf{a}\rangle = \hbar |\omega_{+},\mathbf{a}\rangle$ where $\hat{L}_{z,\mathbf{a}} = (\hat{\mathbf{r}}-\mathbf{a})\times \hat{\mathbf{p}}$ is the angular momentum about the center $\mathbf{a}$.

Consider the group elements $T_{\mathbf{r}}$ and $R_{\mathbf{r}}^{\theta}$ of Euclidean Group E(2) representing translation and rotation by $\theta$ about $\mathbf{r}$ respectively. Their corresponding operators can be expressed in terms of momentum and angular momentum as $\hat{T}_{\mathbf{r}}=\exp(-\im \hat{\mathbf{p}}\cdot\mathbf{r}/\hbar)$ and $\hat{R}_{\mathbf{r}}^{\theta} = \exp(-\im \theta\hat{L}_{z,\mathbf{r}}/\hbar)$. Geometrically, these operators are related as $\hat{R}_{\mathbf{a}}^{\theta} = \hat{T}_{\mathbf{a}}\hat{R}_{0}^{\theta}\hat{T}_{\mathbf{a}}^{\dagger} = \hat{T}_{\mathbf{a}-R_{0}^{\theta}\mathbf{a}}\hat{R}_{0}^{\theta} $~\cite{tung1985group}. 

Under translations and rotations, the Wannier states transform as $\hat{T}_{\mathbf{r}}|\omega_{+},\mathbf{a}\rangle = |\omega_{+},\mathbf{a}+\mathbf{r}\rangle$ and $\hat{R}_{\mathbf{a}}^{\theta}|\omega_{+},\mathbf{a}\rangle = \exp(-\im \theta)|\omega_{+},\mathbf{a}\rangle$ 

Using these relations, we can now prove Eq.~\eqref{eqn:sym} (where $\hat{R}$ is a shorthand for $\hat{R}^{\theta}_{0}$) through a direct computation: 
\begin{align}
    \hat{R}_{0}^{\theta}|\psi_{+}(\mathbf{K})\rangle &= \hat{R}_{0}^{\theta}  \sum_{\mathbf{a} \in \Lambda} \exp(\im \mathbf{K}\cdot\mathbf{a})|\omega_+,\mathbf{a}\rangle \notag \\
    &= \sum_{\mathbf{a} \in \Lambda} \exp(\im \mathbf{K}\cdot\mathbf{a}) \hat{T}_{R_{0}^{\theta}\mathbf{a}-\mathbf{a}}\hat{R}_{\mathbf{a}}^{\theta} |\omega_+,\mathbf{a}\rangle \notag \\
    &= \exp(-\im \theta) \sum_{\mathbf{a} \in \Lambda} \exp(\im \mathbf{K}\cdot\mathbf{a}) \hat{T}_{R_{0}^{\theta}\mathbf{a}-\mathbf{a}} |\omega_+,\mathbf{a}\rangle \notag \\
    &= \exp(-\im \theta) \sum_{\mathbf{a} \in \Lambda} \exp(\im \mathbf{K}\cdot\mathbf{a}) |\omega_+,R_{0}^{\theta}\mathbf{a}\rangle \notag \\
    &= \enspace \exp(-\im \theta) \sum_{\mathbf{a} \in \Lambda} \exp(\im R_{0}^{\theta}\mathbf{K}\cdot R_{0}^{\theta}\mathbf{a}) |\omega_+,R_{0}^{\theta}\mathbf{a}\rangle \notag \\
    &= \enspace \exp(-\im \theta) \sum_{R_{0}^{\theta}\mathbf{a} \in \Lambda} \exp(\im R_{0}^{\theta}\mathbf{K}\cdot R_{0}^{\theta}\mathbf{a}) |\omega_+,R_{0}^{\theta}\mathbf{a}\rangle \notag \\
    &= \exp(-\im \theta) |\psi_{+}(R_{0}^{\theta}\mathbf{K})\rangle \notag \\
    &= \ \exp(-\im \theta) |\psi_{+}(\mathbf{K})\rangle \label{eqn:proof}
\end{align}

where in the first line, we use $\Lambda$ to represent our lattice sites. In the third and fourth line, we use the translation and rotation properties of the Wannier states. In the fixth line, we employ the expression $\mathbf{K}\cdot\mathbf{a}=R_{0}^{\theta}\mathbf{K}\cdot R_{0}^{\theta}\mathbf{a}$. In the sixth line, we use the fact that $R_{0}^{\theta}$ is the symmetry of lattice $\Lambda$, which means that$R_{0}^{\theta}\mathbf{a}$ is a lattice vector as long as $\mathbf{a}$ is a lattice vector. In the last line, we utilize the fact that $R_{0}^{\theta}\mathbf{a}$ lies within the little group of $\mathbf{K}$ (Fig.~\ref{fig:BZ}) , which implies that $R_{0}^{\theta}\mathbf{K}$ is equivalent to $\mathbf{K}$ up to some reciprocal lattice $\mathbf{b}$. Consequently, we have $|\psi_{+}(R_{0}^{\theta}\mathbf{K})\rangle = |\psi_{+}(\mathbf{K})\rangle$.

Actually we can see that the proof holds even if the Wannier state lacks continuous rotational symmetry but possesses discrete rotation symmetry $\hat{R}_{\mathbf{a}}^{\theta}|\omega_{+},\mathbf{a}\rangle = \exp(-\im \theta)|\omega_{+},\mathbf{a}\rangle$ instead.

Additionally, it is worth noticing that the proof presented above is not restricted to a particular high symmetry point in the Brillouin Zone, but rather applies to any high symmetry point with non-trivial little group. 
Let's consider chiral $p$-BEC experiments on a hexagonal lattice as an example. In the experiments described in~\cite{wang2021evidence}, atoms will spontaneously break TRS and condense at either $|\psi_{+}(\mathbf{K}_{\triangle})\rangle^{\otimes N}$ or $|\psi_{-}(\mathbf{K}_{\nabla})\rangle^{\otimes N}$. Here, the little groups of high symmetry points $\mathbf{K}_{\triangle,\nabla}$ are $C_3$ generated by a rotation of $\theta_{K} = 2\pi/3$ about the origin. For such states, we can substitute $\theta$ with $\theta_K = 2\pi/3$ in Eq.~\eqref{eqn:proof} and utilize the same approach to establish $\hat{R}_{0}^{2\pi/3}|\psi_{+}(\mathbf{K}_{\triangle})\rangle = \exp(-\im 2\pi/3)|\psi_{+}(\mathbf{K}_{\triangle})\rangle$. 
In other parameter regime in hexagonal lattice experiments~\cite{jin2021evidence,wang2022observation}, instead of the $\mathbf{K}_{\triangle,\nabla}$ points, the chiral $p$-BEC state may condense at the $M_{1,2,3}$ points whose little group is $C_2$ with $\theta_M = \pi$. 
Moreover, We can extend this result to other lattices. For example, the $p$-BEC state in checkerboard lattice condenses at $\mathbf{K}_{(\pm,\pm)}$, which are located at the midpoints of corners of square-shaped first Brillouin Zone. These high symmetry points $\mathbf{K}_{(\pm,\pm)}$ also possess a little group of $C_2$ symmetry~\cite{wirth2011evidence}.

\newpage

\section{S2. Scattering of LLL condensate state under the contact interaction}
The interacting many body hamiltonian~\eqref{eqn:hamiltonian} has some well-known properties~\cite{GSexpression,DoAttractive,lewin2009strongly,FqheExact}:

In this section, we prove the fact that in our system when $g \ll \hbar \omega_{xy,z}$ and $\hat{V}_{\text{int}}$ can be treated as a perturbation, LLL states will always be scattered into other LLL states rather than higher LL states. The full many-body Hilbert space for $N$ bosonic particles in a 2d harmonic trap is spanned by states of the form $\mathcal{S}\left(\bigotimes_{i=1}^{N}|n_i,m_i\rangle\right)$ with $(n_i,m_i)$ the energy and angular momentum quantum number of the $i$th particle, and $\mathcal{S}$ the symmetrization operator to enforce bosonic statistics. Suppose an initial state $\mathcal{S}\left(\bigotimes_{i=1}^{N}|n_i,m_i\rangle\right)$ is scattered into a final state $\mathcal{S}\left(\bigotimes_{i=1}^{N}|n'_i,m'_i\rangle\right)$ due to the perturbative interaction $\hat{V}_{\text{int}}$. The conservation of energy and angular momentum dictates that
\begin{align}
    \sum_{i=1}^N n_i & = \sum_{i=1}^N n'_i \label{eqn:conservation of energy} \\ 
    \sum_{i=1}^N m_i &= \sum_{i=1}^N m'_i \label{eqn:conservation of angular momentum}
\end{align}

For LLL initial states, $m_i=n_i$ and therefore the scattered states satisfy $\sum_{i=1}^N m'_i = \sum_{i=1}^N n'_i$. Since the range of $m$ is $n, n-2, ..., -n$, we have $m'_i \leq n'_i$ and
\begin{equation}\label{eqn:inequality}
     \sum_{i=1}^N n'_i = \sum_{i=1}^N m'_i \leq \sum_{i=1}^N n'_i
\end{equation}

The equality in Eq.\eqref{eqn:inequality} holds true only if $m'_i = n'_i, \ i=1,2,..,N$. This condition implies that the scattered states still belong to the LLL.

\section{S3. The first excited state in \texorpdfstring{$\mathcal{H}_{N,L=N}$}{Lg}}
\label{sec:appendixB}
In Fig.~\ref{fig:gap} we presented numerical evidence that $\hat{V}_{\text{int}}=\sum_{i<j} \hat{V}_{ij} = 2\pi g\sum_{i<j} \delta(\mathbf{r}_i-\mathbf{r}_j)$ has $\epsilon_0 = gN(N-2)/4$, $\epsilon_1 = gN(N-1)/4$ and $\Delta = gN/4$ in $\mathcal{H}_{N,L=N}$ when $N\geq 10$. Actually a rigorous proof of the ground state energy $\epsilon_0$ and the exact wavefunction of the ground state $\psi_0$ in $\mathcal{H}_{N,L\leq N}$ has been provided in~\cite{lewin2009strongly,GSexpression}. The ground state wavefunction $\psi_0$ is given by

\begin{equation}
    \psi_0^{N,L}(z_1,...,z_N) = \sum_{1 \leq p_1 < p_2 < ... < p_L \leq N} \prod_{i=1}^{L} \left[(z_{p_i}-z_c)\cdot\e^{-|z_i|^2/2}\right]
\end{equation}
where $z_c = (z_1+...+z_N)/N$ represents the center of mass coordinate. The ground state energy corresponding to $\psi_0^{N,L}$ is 
\begin{equation}
    \epsilon_0^{N,L} = \frac{g}{2}\left(N(N-1)-\frac{NL}{2}\right)
\end{equation}

Now in this section, we will show the existence of an eigenstate $\psi_1$ of $\hat{V}_{\text{int}}$ with energy $\epsilon_1$, though we cannot rigorously prove that $\psi_1$ is the first excited state of $\hat{V}_{\text{int}}$ .

By omitting the common factor $\exp(-1/2\sum|z_i|^2)$ of wavefunctions in $\mathcal{H}_{N,L}$, we obtain the Bargmann space $\mathcal{B}_{N,L}$ which consists of totally symmetric, analytic and homogeneous polynomial $F(z_1,...,z_N)$ of degree $L$~\cite{QHEformalisim,GSexpression,LiebPRA2009}. In this space, the angular momentum and contact interaction act as $\hat{L}_i = z_i\partial_i$ and $\hat{V}_{ij} = \hat{\delta}_{ij}$ respectively. Here $\hat{\delta}_{ij}$ acts on the functions as $\hat{\delta}_{ij} F(z_1,...,z_i,...,z_j,...,z_N) = F(z_1,...,(z_i+z_j)/2,...,(z_i+z_j)/2,...,z_N)$. Two important properties of $\hat{\delta}_{ij}$ are as follows: (a) The center of mass coordinate $z_c = (z_1+...+z_N)/N$ is invariant under $\hat{\delta}_{ij}$ as $\hat{\delta}_{ij}z_c = z_c$. (b) $\hat{\delta}_{ij}$ preserves the composition of functions, meaning that for all $ F,G \in \mathcal{B}_N$, we have $\hat{\delta}_{ij}(F\cdot G) = (\hat{\delta}_{ij}F) \cdot (\hat{\delta}_{ij}G)$.

Consider the spurious state $\psi_{1}^{N,L=N}$, which is constructed by exciting the center of mass coordinate in the ground state of the subspace $\mathcal{B}_{N,L=N-1}$ :
\begin{equation}
    \psi_{1}^{N,L=N} = z_c \cdot \psi_{0}^{N,L=N-1} = z_c \sum_{1 \leq p_1 < p_2 < ... < p_{N-1} \leq N} \prod_{i=1}^{N-1} (z_{p_i}-z_c) \in \mathcal{B}_{N,L=N}
\end{equation}

It can be shown that $\psi_{1}^{N,L=N}$ is an exact eigenstate with $\epsilon_1 = gN(N-1)/4 = \epsilon_0^{N,N-1}$ because
\begin{align}
    \hat{V}_{\text{int}}\psi_{1}^{N,L=N} &= \sum_{1\leq i<j\leq N} \hat{\delta}_{ij} \ \left(z_c \cdot \psi_0^{N,L=N-1} \right) \notag \\
    &= \sum_{1\leq i<j\leq N} \Big(\hat{\delta}_{ij} \ z_c\Big) \cdot \Big(\hat{\delta}_{ij} \ \psi_{0}^{N,L=N-1}\Big) \notag \\
    &= \sum_{1\leq i<j\leq N} z_c \cdot \Big(\hat{\delta}_{ij} \ \psi_{0}^{N,L=N-1}\Big) \notag \\
    &= z_c \cdot \Big( \hat{V}_{\text{int}} \ \psi_0^{N,L=N-1} \Big) \notag \\
    &= z_c \cdot \Big( \epsilon_0^{N,L=N-1} \ \psi_0^{N,L=N-1} \Big) = \epsilon_1 \psi_{1}^{N,L=N}
\end{align}

where in the second line we use property (b): $\hat{\delta}_{ij}$ preserves the composition of $z_c$ and $\psi_0^{N,L=N-1}$, and in the third line we use the property (a): $\hat{\delta}_{ij}z_c = z_c$.

\section{S4. $p$-BEC states in hexagonal and triangular optical lattice}
In our main text, we adopt the experimental setup introduced in~\cite{wang2021evidence,jin2021evidence}, which realizes a $p$-BEC in hexagonal optical lattice. The hexagonal optical lattice consists of two triangular optical lattice with sites denoted as $\mathcal{A}$ and $\mathcal{B}$, corresponding to the two local potential minimum. The hexagonal lattice is generated by three laser beams with wave vectors $\mathbf{k}_1 = (-1,0)k_L,\ \mathbf{k}_{2,3} = (1/2,\pm\sqrt{3}/2)k_L$. The lattice potential takes the form
\begin{equation}\label{eqn:hexagonal}
    V_{\text{hex}} = -2V_0\sum_{i=1}^{3}\cos(\mathbf{k}_i\cdot \mathbf{r}) -2V_1\sum_{i=1}^{3}\cos(\mathbf{k}_i\cdot \mathbf{r} +\phi_i) \text{  with  } (\phi_1,\phi_2,\phi_3) = (4\pi/3,2\pi/3,0)
\end{equation}
where the first(second) summation term describes the triangular optical lattice with $\mathcal{A}$($\mathcal{B}$) sites.

\begin{figure}[h]
\includegraphics[width=0.85\linewidth]{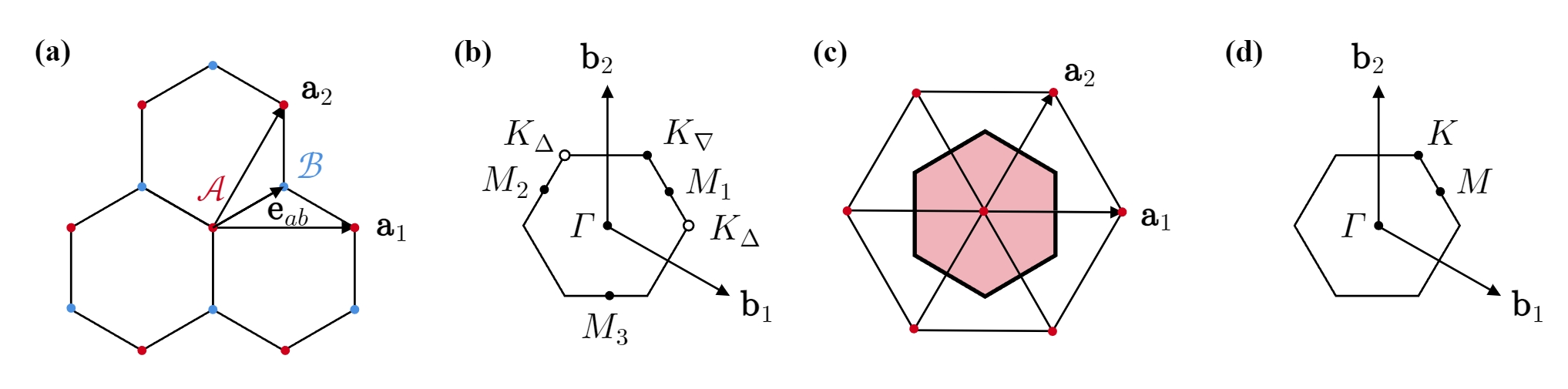}
\caption{Hexagonal optical lattice, triangular optical lattice and their common first Brillouin Zone.
(\tbf{a}-\tbf{b}) The hexagonal optical lattice and triangular optical lattice share common lattice vectors:$\mathbf{a}_{1}=a(1,0),\mathbf{a}_2=a(1/2,\sqrt{3}/2)$ where the lattice constant $a=4\pi/3k_L$.
(\tbf{a}) Hexagonal optical lattice $V_{\text{hex}}$ with  $\mathcal{A}$ and $\mathcal{B}$ sites described by \eqref{eqn:hexagonal}. $\mathbf{e}_{ab} = a(1/2,\sqrt{3}/6)$ connects nearest-neighbor $\mathcal{A}$ and $\mathcal{B}$ sites.
(\tbf{b}) The first Brillouin Zone of hexagonal with high symmetry points $\Gamma,M,K$. The reciprocal lattice vectors are $\mathbf{b}_1=\sqrt{3}k_L(\sqrt{3}/2,-1/2$ and $\mathbf{b}_2=\sqrt{3}k_L(0,1)$. There are two inequivalent $K$ points, labeled as $K_{\triangle}$ and $K_{\nabla}$. The little group of these two $K$ points is $C_3$, generated by a rotation of $\theta=2\pi/3$ around the origin $\Gamma$. There are also three inequivalent $M$ points, labeled as $M_{1,2,3}$.
(\tbf{c}) Triangular optical lattice $V_{\text{lat}}$ with only $\mathcal{A}$ sites described by \eqref{eqn:triangular}.  The central pink unit cell is the Wigner-Seitz unit cell of the lattice, which is the only unit cell considered in our adiabatic protocol.
(\tbf{d}) The first Brillouin Zone of hexagonal with high symmetry points $\Gamma,M,K$. The $K,K'$ and $M_{1,2,3}$ are now equivalent high symmetry point. Their little groups are all $C_6$.
}
\label{fig:BZ}
\end{figure}

In these experiments,by tuning the relative depth of $\mathcal{A}$ and $\mathcal{B}$ sites, the atoms are loaded into the second Bloch band which has 2 degenerate energy minima at the K points, which are denoted as $\mathbf{K}_{\triangle}$ and $\mathbf{K}_{\nabla}$. Their degeneracy is protected by TRS. In the weak interaction limit, the atoms will spontaneously break TRS and condense at either $|\psi_{+}(\mathbf{K}_{\triangle})\rangle^{\otimes N}$ or $|\psi_{-}(\mathbf{K}_{\nabla})\rangle^{\otimes N}$. The chiral $p$-BEC state $|\psi_{+}(\mathbf{K}_\triangle)$ can be expressed as a superposition of $p$-orbital Wannier state localized at $\mathcal{A}$ site and $s$-orbital Wannier state localized at $\mathcal{B}$ site as
\begin{equation}\label{eqn:pHexagonal}
    |\psi_{+,\text{hex}}(\mathbf{K}_{\triangle})\rangle = \sum_{\mathbf{R}_a\in \mathcal{A}} \exp (i\mathbf{K}_{\triangle}\cdot\mathbf{R}_a) \big[C_a|\omega_{+},\mathbf{R}_a\rangle + C_b|\omega_{s},\mathbf{R}_b\rangle  \big]
\end{equation}
where $\mathbf{R}_b = \mathbf{R}_a +\mathbf{e}_{ab} \in \mathcal{B}$ represents the location of $\mathcal{B}$ sites. $\mathbf{e}_{ab}$ is the vector connecting nearest-neighbor $\mathcal{A}$ and $\mathcal{B}$ sites(see Fig.~ \ref{fig:BZ}(\tbf{a})), and $C_{a,b}$ are coefficients determined by the depth $V_0,V_1$ of lattice potential.

As calculated in~\cite{wang2021evidence}, it was found that by decreasing the value of $V_1$ and making the $\mathcal{B}$ wells shallower, the population in the $p$-orbital increases. Finally, if the triangular optical lattice with $\mathcal{B}$ sites is completely turned off, all states will be transferred to the $p$-orbitals localized at $\mathcal{A}$ sites.

After switching off the triangular optical lattice with $\mathcal{B}$ sites, the hexagonal optical lattice described by \eqref{eqn:hexagonal} now degenerates to a triangular optical lattice with only $\mathcal{A}$ sites used in our example
\begin{equation}\label{eqn:triangular}
    V_{\text{lat}} (\mathbf{r}) = -2V_0\sum_{i=1}^{3}\cos(\mathbf{k}_i\cdot \mathbf{r})
\end{equation}

There is no contribution from the $s$-orbitals localized at the original $\mathcal{B}$ sites now. Therefore, the $p$-BEC state $|\psi_{+}(\mathbf{K}_\triangle)$ can be expanded as
\begin{equation}\label{eqn:pTriangular}
    |\psi_{+}(\mathbf{K})\rangle = \sum_{\mathbf{R}_a\in \mathcal{A}} \exp (i\mathbf{K}\cdot\mathbf{R}_a) |\omega_{+},\mathbf{R}_a\rangle
\end{equation}

\section{S5. Evolution of the gap in the full Adiabatic Procedure}
In this section, we provide a detailed demonstration of how the gap in our full adiabatic procedure evolves. Throughout the adiabatic process, we set the interaction strength $g$ between atoms in Eq.~\eqref{eqn:hamiltonian} to be $0$, thereby decoupling the evolution of the $xy$ and $z$ sectors from each other. Specifically,
\begin{equation}
    |\psi(t)\rangle = |\psi_{xy}(t)\rangle \otimes |\psi_z(t)\rangle  \text{  with } |\psi_\alpha(t)\rangle = \exp(-\im \hat{H}_\alpha t/\hbar) \ ,\alpha = xy  \text{ or } z
\end{equation}

where $\hat{H}_{xy}(t) = (\hat{p}_x^2+\hat{p}_y^2)/2M + V_{xy}(t)$ and $\hat{H}_{z}(t) = \hat{p}_z^2/2M + V_{z}(t)$ are the single body hamiltonians of $xy$ and $z$ sector. The potential field in the full protocol are summarized in Table. \ref{table:1}.

\begin{table}[h]
\centering
\begin{tabular}{ | M{2em} | M{4cm}| M{4cm} | } 
 \hline
  & \textbf{Step I} : $0\leq t\leq T_1$ &\textbf{Step II} : $T_1\leq t\leq T_1+T_2$ \\ 
 \hline
 $V_{xy}(t)$ & $(1-t/T_1)V_{\text{lat}} + t/T_1\cdot V_{\text{trap,xy}}$ & $\frac{1}{2}M \omega_{xy}^2(t) (x^2+y^2)$\\ 
 \hline
 $V_{z}(t)$ & $\frac{1}{2}M \omega_{z}^2(0) z^2$ & $\frac{1}{2}M \omega_{z}^2(t) z^2$ \\ 
 \hline
\end{tabular}
\caption{Potential field in the full adiabatic procedure.}
\label{table:1}
\end{table}

As a result, we can examine the gaps in two sectors separately. The results are listed as follows:

\begin{table}[h]
\centering
\begin{tabular}{ | M{6em} | M{4cm}| M{4cm} | } 
 \hline
  Gap in $\mathcal{H}_{\text{sym}}$ & \textbf{Step I} : $0\leq t\leq T_1$ &\textbf{Step II} : $T_1\leq t\leq T_1+T_2$ \\ 
 \hline
 $\Delta_{xy}(t)$ & shown in Fig.~\ref{fig:energy_level} & 2$\hbar\omega_{xy}(t)$\\ 
 \hline
 $\Delta_{z}(t)$ & $2\hbar\omega_{z}(0)$ & $2\hbar\omega_z(t)$ \\ 
 \hline
\end{tabular}
\caption{Evolution of the gap in $\mathcal{H}_{\text{sym}}$ in the full adiabatic procedure}
\label{table:2}
\end{table}

In the second step, the hamiltonian in both $xy$ and $z$ sectors are harmonic oscillators and our state remains as $|\psi(t)\rangle = |n=1,m=1\rangle \otimes |n_z=0\rangle$. In the full Hilbert space without any symmetry restriction, $|n=1,m=\pm1\rangle$ are degenerate with $\hat{H}_{xy}$ and the gap between $|n=1,m=\pm1\rangle$ and higher energy states in the $xy$ sector is $\hbar \omega_{xy}$. In the $z$ sector, the gap $\Delta(z)(t)=\hbar \omega_z(t)$ is the energy difference between $|n_z=0\rangle$ and $|n_z = 1\rangle$.

Since discrete lattice rotation $\hat{R}=\exp(-\im \theta \hat{L}_z)=\exp(-\im \theta)$ and parity in z $P_z : z \rightarrow -z,\ \hat{P}_z = 1$ are conserved during the adiabatic evolution, these two symmetries force $\psi(t)$ to evolve in the symmetrical subspace $\mathcal{H}_{\text{sym}}$, subject to the restrictions $\hat{R}=\exp(-\im \theta)$ and $\hat{P}_z = 1$. Within this subspace, $\hat{R}=\exp(-\im \pi/3 \hat{L}_z)=\exp(-\im \pi/3)$ enforces $\psi_{xy}(t)$ to be a superposition of states with $\hat{L}_z=m=1+6a, a\in \mathbb{Z}$, while $\hat{P}_z = 1$ enforces $\psi_z(t)$ to contain states with only $n_z \in 2\mathbb{Z}$. Consequently, $|n=1,m=1\rangle$ and $|n=3,m=1\rangle$ are the ground state and first excited state in the $xy$ sector, and $|n_z=0\rangle$ and $|n_z=2\rangle$ are the ground state and first excited state in the $z$ sector. Therefore, $\Delta_{xy}(t) = \hbar \omega_{xy}(t), \Delta_z(t) = 2\hbar \omega_z(t)$ and the full gap $\Delta(t)=\min(2\hbar\omega_{xy}(t),2\hbar \omega_{z}(t))$. For the same reason, $\Delta_z(t)=2\hbar \omega_z(0)$ in the first step. $\Delta_{xy}(t)$ in the first step is calculated numerically by using exact diagonalization method. The evolution of the gap in two sectors in the full adiabatic procedure is summarized in Table.~\ref{table:2}. The application of adiabatic theorem in the two sectors ensure the feasibility of our protocol.

\end{document}